\documentclass[11pt,showpacs]{revtex4-1}

\usepackage{epsfig}
\input epsf.tex
\usepackage{amssymb}
\usepackage{amsmath}
\usepackage{amsthm}
\usepackage{amsopn}

\def\comment#1{}

\def\beq{\begin{equation}}
\def\eeq{\end{equation}}
\def\bea{\begin{eqnarray}}
\def\eea{\end{eqnarray}}

\usepackage{ulem}
\usepackage{cancel}
\usepackage{color}

\begin{document}

\title{ Photon-neutrino scattering and the B-mode spectrum of CMB photons}

\author{  Jafar Khodagholizadeh$^{*}$, Rohoollah Mohammadi$^{*\dag}$\footnote{rmohammadi@ipm.ir} and She-Sheng Xue$^{\ddag}$\footnote{xue@icra.it}}
\affiliation{$^{*}$School of physics, Institute for research in fundamental sciences (IPM), Tehran, Iran.\\
$^{\dag}$Iran Science and Technology Museum   (IRSTM), PO BOX: 11369-14611, Tehran, Iran.\\
$^{\ddag}$ICRANet, P.zza della Repubblica 10, I--65122 Pescara, Physics Department, University of Rome {\it La Sapienza}, P.le Aldo Moro 5, I--00185 Rome, Italy.}

\begin{abstract}
On the basis of the quantum Boltzmann equation governing the time-evolution
of the density matrix of polarized CMB photons in the primordial scalar
perturbations of metric, we calculate the B-mode spectrum of polarized CMB
photons contributed from the scattering of CMB photons and CNB neutrinos
(Cosmic Neutrino Background). We
show that such contribution to the B-mode spectrum is
negligible for small $\ell$, however is significantly large for $50 < \ell< 200$ by plotting our results together with the BICEP2 data. Our study and results imply that in order to theoretically better understand the origin of the observed B-mode spectrum of polarized CMB
photons ($r$-parameter), it should be necessary to study the relevant and dominate processes in both tensor and scalar perturbations.
\end{abstract}

\pacs{13.15.+g, 98.80.Es, 98.70.Vc.}

\maketitle

\section{Introduction}

It is known that in the inflation cosmology, the power-law spectrum
of either metric scalar
perturbation $P_{S} (k)=A_{S}(k/k_{0})^{n_{S}-1}$
or tensor perturbation $P_{T} (k)=A_{T}(k/k_{0})^{n_{T}-1}$
 has been produced in the inflationary era of the early universe \cite{Lidsey},
where $A_{S}$ and $A_{T}$ are the amplitudes of
scalar and tensor perturbations, and $n_{S,T}$ are their spectral indexes.
$A_{S}$ and $n_{S}$ have been determined through the measurements of microwave
background temperature anisotropy \cite{Sievers,Hou,Ade}.
The amplitude of metric
tensor perturbation is characterized by the tensor-scalar ratio
$r=P_{T}/P_{S}$, relating to the B-mode spectrum of polarized CMB photons
imprinted by the metric tensor perturbations of
primordial gravitational waves.
The BICEP2 collaboration recently reports $r=0.20_{-0.05}^{+0.07}$
\cite{BICEP2}. If this report is verified, it is regarded as an
important result that may reveal the existence of metric tensor
perturbations in the inflationary era of the early universe.

However, there are alternative explanations of the BICEP2 data.  Whether the
BICEP2 data could be explained by the vector and tensor modes from primordial
magnetic fields \cite{Bovin}. Some authors speculate that the BICEP2
observed B-mode polarization is the result of a primordial Faraday rotation
of the E-mode polarization \cite{14,Giovannini}. In this article, using the
result of photon polarization generated
by the photon-neutrino scattering \cite{roh}, we
investigate the possible contribution to the observed B-mode spectrum by
considering the interaction between CMB photons and Cosmic Neutrino
Background (CNB) in the background of scalar perturbations, without tensor
perturbations. In order to quantitatively calculate such contribution in the
scalar perturbation,
we solve the quantum Boltzmann equation for the time-evolution of the matrix
density (Stokes parameters) of polarized CMB photons  which are involved in the
Compton and photon-neutrino scatterings as the collision terms of the quantum
Boltzmann equation. Our result is shown together with the BICEP2 data and its implication on the interpretation of the BICEP2 data is discussed.

\section{The photon polarization from Compton and photon-neutrino scatterings.}

The linear and circular polarizations of an ensemble of photons can be described by the density operator
\bea
\hat\rho_{ij}=\frac{1}{\rm {tr}(\hat \rho)}\int\frac{d^3k}{(2\pi)^3}
\rho_{ij}(k)D_{ij}(k),\quad \hat{\rho}_{ij}(k)=\frac{1}{2}\left(\begin{array}{cc}
             I+Q& U-iV \\
             U+iV & I-Q \\
                 \end{array}
        \right),\label{t0}
\eea
where $\rho_{ij}(k)$ represents the density matrix in terms of the
Stokes parameters I, Q, U and V in the $2\times 2$ polarization space
($i,j$) of one photon of energy-momentum ``$k$''.
The number operator $D_{ij}(k)=a^\dagger_{i}(k)a_{j}(k)$ and its
expectation value
\bea
\langle\, D_{ij}(k)\,\rangle\equiv {\rm tr}[\hat\rho_{ij}
D_{ij}(k)]=(2\pi)^3 \delta^3(0)(2k^0)\rho_{ij}(k).\label{t1}
\eea
The time evolution of the number operator $D_{ij}(k)$ obeys the Heisenberg equation
\begin{equation}\label{heisen}
   \frac{d}{dt} D_{ij}(k)= i[H_I,D_{ij}(k)],
\end{equation}
where $H_I$ is an interacting Hamiltonian. Using Eqs.~(\ref{t0}), (\ref{t1})
and (\ref{heisen}), one obtains the time evolution of $\rho_{ij}(k)$,  quantum Boltzmann equation \cite{cosowsky1994},
\bea
(2\pi)^3 \delta^3(0)(2k^0)
\frac{d\rho_{ij}(k)}{dt} = i\langle \left[H_I
(t),D_{ij}(k)\right]\rangle-\frac{1}{2}\int dt\langle
\left[H_I(t),\left[H_I
(0),D_{ij}(k)\right]\right]\label{bo}\rangle,
\eea
where $H_I(t)$ is the interacting Hamiltonian. On the right-handed side of
Eq.~(\ref{bo}), the first and second terms respectively represent
the forward scattering and higher order collision terms.
\comment{ whereas
the left-handed side is known as the Liouville term, which deals with the
effects of gravitational perturbations about the homogeneous cosmology.
}

There are a lot of papers which investigate the effects of the Compton scattering on the anisotropy and polarization of CMB (see for example
\cite{cosowsky1994,zal,hu}). In this article, we attempt to study the CMB photon polarization by considering the contribution of the photon-neutrino scattering to the polarization
density matrix of photons obtained recently \cite{roh}
\begin{eqnarray}
  2k^0\frac{d\rho_{ij}}{dt} &=& -\frac{\sqrt{2}}{6\pi}\alpha\,G^F\int d\mathbf{q} \big[\rho_{s'j}({\bf k})\delta_{is} -\rho_{is}({\bf k})\delta_{js'}\big]f_\nu(x,q)\nonumber\\
  &\times& \left(q^2\epsilon_{s'}\cdot\epsilon_s
   +2 \,\mathbf{q}\cdot\epsilon_{s'}\,\,\,\mathbf{q}\cdot\epsilon_s\,-
   \varepsilon_{\mu\nu\rho\sigma}\,\epsilon_s^{\mu}\epsilon_{s'}^\nu k^\rho q^\sigma\right),\label{fws2}
\end{eqnarray}
where $d\mathbf{q}=(2 E_\nu)^{-1}d^3q/(2\pi)^3$ is the integration over the neutrino four-momentum ($q^0=E_\nu \approx |{\bf q}|$) with
the distribution function $f_\nu(x,q)$,
the polarization
four-vectors $\epsilon _{i\mu}({\bf k})$ and their indexes $i,j,s, s'=1,2$,
represent two transverse polarizations of the photon $k^0=|{\bf{k}}|$.  $G^F$
and $\alpha$ are Fermi coupling constant and electromagnetic fine-structure
constant.

Using the Stokes parameters in Eq.~(\ref{t0}):
the total intensity $I$, linear
polarizations intensities $Q$ and $U$, as well as the $V$ indicating
the difference between left- and right-circular polarizations
intensities, we consider the both Compton and photon-neutrino scattering and write Eq.~(\ref{bo}) as follows
\begin{eqnarray}
\frac{dI}{dt}
&=&C^I_{e\gamma} \nonumber \\
\frac{d}{dt}(Q\pm iU)  &=& C^\pm_{e\gamma} \mp i \dot{\kappa}_{\pm} (Q\pm iU)+{\mathcal O}(V)
 \nonumber \\
\frac{dV}{dt}  &=& C^V_{e\gamma} +\dot{\kappa}_{Q}Q + \dot{\kappa}_{U}U,
\label{Bo1}
\end{eqnarray}
here $C^I_{e\gamma}$, $C^\pm_{e\gamma}$ and $C^V_{e\gamma}$ respectively
indicate the contributions from the Compton scattering to the time
evaluation of $I$, $Q\pm iU$ and $V$ parameters, their expressions can be found from the literature for example \cite{cosowsky1994,zal,hu}. Whereas the contributions
from the photon-neutrino scattering (\ref{fws2}) are given by
\begin{eqnarray}
  \dot{\kappa}_{\pm}&=&  -\frac{\sqrt{2}}{6\pi k^0}\alpha\,G^F\int d\mathbf{q}\, f_\nu(x,q)
  \times \left(
   \varepsilon_{\mu\nu\rho\sigma}\,\epsilon_2^{\mu}\epsilon_{1}^\nu k^\rho q^\sigma\right)\nonumber\\
  \dot{\kappa}_{Q} &=& -\frac{\sqrt{2}}{3\pi k^0}\alpha\,G^F\,n_\nu
   \,\langle  v_\alpha q_\beta \rangle\epsilon^\alpha_{2}\,\epsilon^\beta_1\nonumber \\
  \dot{\kappa}_{U} &=&  -\frac{\sqrt{2}}{6\pi k^0}\alpha\,G^F\,n_\nu\,\left(
    \,\langle  v_\alpha q_\beta \rangle\epsilon^\alpha_{1}\,\epsilon^\beta_1-\,\langle  v_\alpha q_\beta \rangle\epsilon^\alpha_{2}\,\epsilon^\beta_2\,\right),\label{cBo1}
\end{eqnarray}
where we define the neutrino average velocity
\begin{eqnarray}
  \langle v_\alpha  \rangle = \frac{1}{n_\nu}\int\frac{d^3q}{(2\pi)^3}\,\frac{q_\alpha}{q_0}\,f_\nu(x,q), \quad
  \langle  v_\alpha q_\beta \rangle = \frac{1}{n_\nu}\int\frac{d^3q}{(2\pi)^3}\,\frac{q_\alpha}{q_0}\,q_\beta\,f_\nu(x,q). \label{average}
\end{eqnarray}
where the neutrino number-density $n_\nu(x)=\int d^3q/(2\pi)^3 f_\nu(x,q)$ and
energy-density $\epsilon_\nu(x)=\int d^3q/(2\pi)^3 q^0f_\nu(x,q)$.
In the second equation of Eq.~(\ref{Bo1}), we will neglect the small
contribution ${\mathcal O}(V)$ from the circular polarization.
In Eq. (\ref{cBo1}), the first equation $\dot{\kappa}_{\pm}$ yields
\begin{eqnarray}
\dot{\kappa}_{\pm} &=&  \frac{\sqrt{2}}{6\pi k^0}\alpha\,G^F\int d\mathbf{q}\, f_\nu(x,q)
  \times \left[q^0 {\bf k}\cdot(\epsilon_1\times\epsilon_2)+k^0 {\bf q}\cdot(\epsilon_1\times\epsilon_2)\right]\nonumber\\
&=&  \frac{\sqrt{2}}{6\pi}\alpha\,G^F \frac{n_\nu}{2}\left[1+\langle {\bf v}\rangle\cdot(\epsilon_1\times\epsilon_2)\right]\approx \frac{\sqrt{2}}{6\pi}\alpha\,G^F \frac{n_\nu}{2},\label{kappa1}
\end{eqnarray}
where ${\bf k}\cdot(\epsilon_1\times\epsilon_2)= |{\bf k}|$. In this article, we apply these equations (\ref{Bo1}\,-\,\ref{kappa1}) to the case of photons scattering with cosmic neutrino background (CNB), whose
average velocities (\ref{average}) are small, will be discussed in the next section. As a result of the leading order approximation, the dominate contribution of photon-neutrino scattering to photon polarization comes from the first term of Eq.(\ref{kappa1}).

\section{CNB neutrino distribution function and average velocity}\label{int}

We discuss the CNB neutrino distribution function $f_\nu(x,q)$ and average velocity $\langle {\bf v}\rangle$ by using the Boltzmann equation for massive neutrinos (see the Chapter 4 of
Ref.~\cite{MC}).
Because the ensemble of massive neutrinos behaves as a
hydrodynamic fluid described by the distribution function $f_\nu(x,q)$,
satisfying the following Boltzmann evolution equation derived from the
conservations of neutrino number and energy-momentum,
\begin{equation}\label{bonu}
    \frac{\partial f_\nu}{\partial t}+\frac{p_i}{a E_\nu} \frac{\partial f_\nu}{\partial x_i}-
    \frac{\partial f_\nu}{\partial E_\nu}\Big(\frac{\dot a}{a}\frac{p^2}{E_\nu}+\frac{p^2}{E_\nu}\frac{\partial \phi}{\partial t}+\frac{p_i}{a}\frac{\partial \psi}{\partial x_i}\Big)=0,
\end{equation}
where the longitudinal gauge
\cite{mukhanov} is adopted and the scalar modes of metric
perturbations are characterized by two scalar potentials $\phi$ and $\varphi$
in the line element
\begin{equation}\label{line}
    ds^2=a^2(\tau)\{-(1+2\psi)d\tau^2+(1+2\varphi)dx_i\,dx^i\}.
\end{equation}

We are in the comoving frame of expanding universe, where the homogeneous and
isotropic density $\bar n_\nu\propto a^{-3}$ and the neutrino average
velocities $ \langle \bar v^i\rangle$ are zero.
However, we observe the perturbation of the
neutrino fluid density $\delta_\nu\equiv \delta n_\nu /\bar n_\nu= (n_\nu-\bar n_\nu)/\bar n_\nu$ and
the perturbation of the neutrino average velocities
$\delta v^i\equiv (\langle v^i\rangle-\langle \bar v^i\rangle)=\langle v^i\rangle$.
The perturbation of the neutrino average velocities
$\delta \langle v^i\rangle=\langle v^i\rangle$ is proportional to the perturbation of the
neutrino fluid density $\delta_\nu$. This can be seen by the linear approximation of the Boltzmann equation (\ref{bonu}).
Integrating the Boltzmann equation (\ref{bonu}) over
$\int\frac{d^3p}{(2\pi)^3}$,
$\int\frac{d^3p}{(2\pi)^3}\frac{ p_i}{E_\nu}$ and
$\int\frac{d^3p}{(2\pi)^3}\frac{ p_i}{E_\nu}\frac{ p_j }{E_\nu}$, and considering the linear approximation, one obtains
\begin{eqnarray}
  \frac{\partial \delta_\nu}{\partial t}+ \frac{1}{a}\frac{\partial
	\langle v^i\rangle}{\partial x^i}+3\frac{\partial \varphi}{\partial t}&=& 0 \label{dnu1}\\
  \frac{\partial \langle v^i \rangle}{\partial t}+\frac{\dot a}{a}\langle v^i \rangle+\frac{1}{a}\frac{\partial \psi}{\partial x_i}&=& 0\label{velo1}\\
   \frac{\partial \langle v^i v^j \rangle}{\partial t}+2\frac{\dot a}{a}\langle v^i v^j \rangle+\frac{1}{a}\left(\frac{\partial \psi}{\partial x_i}\langle v^j \rangle+\frac{\partial \psi}{\partial x_j}\langle v^i \rangle\right)&=& 0\label{Dvelo1}
\end{eqnarray}
where analogously to Eq.~(\ref{average}) the neutrino average velocities are
\begin{eqnarray}
  \langle v^i\rangle= \frac{1}{n_\nu}\int\frac{d^3p}{(2\pi)^3}\frac{p^i}{E_\nu}f_\nu, \quad
 \langle v^i v^j \rangle = \frac{1}{n_\nu}\int\frac{d^3p}{(2\pi)^3}\frac{p^i}{E_\nu}\frac{p^j}{E_\nu}f_\nu, \label{average1}
\end{eqnarray}
Rewrite Eqs.~(\ref{dnu1})-(\ref{Dvelo1}) in terms of the conformal time
$\eta$ and the Fourier components of variables (e.g.~ $\tilde{v}^i$ for $\langle v^i \rangle$)
\begin{eqnarray}
  \dot{\tilde{\delta}}_\nu+iK\tilde{v}+ 3\dot{\tilde{\varphi}}&=& 0 \label{dnu2}\\
  \dot{\tilde{v}}+\frac{\dot{a}}{a}\tilde{v}+ iK\tilde{\psi} &=& 0 \label{velo2}\\
   \frac{\partial}{\partial \tau}\tilde{v}^i\tilde{v}^j+2\frac{\dot{a}}{a}\tilde{v}^i\tilde{v}^j+ i(K^i\tilde{v}^j+K^j\tilde{v}^i)\tilde{\psi} &=& 0 ,\label{Dvelo2}
\end{eqnarray}
here it is assumed that the velocity is ir-rotational so $v^i=\frac{K^i}{K}\tilde{v}$. As shown in these equations, the neutrino average velocity is of the order of neutrino density perturbations, i.e.,  $\tilde{v}\sim \delta n_\nu/\bar n_\nu \sim \frac{\Delta T}{T}\sim 10^{-5}$. Therefore, in Eq.~(\ref{kappa1}) the second term depending on the neutrino average velocity is negligible, compared with the first term depending on the neutrino density.

In addition, in the following calculations, we select the coordinate where the components of the photon momentum
${\bf k}$, polarization vectors $\epsilon_1$ and $\epsilon_2$ are
\begin{eqnarray}
  k_x &=& \sin\theta\cos\phi,\,\,\,\,\epsilon_{1x}(k)= \cos\theta\cos\phi,\,\,\,\,\epsilon_{2x}(k)= \sin\phi,\nonumber \\
  k_y &=&  \sin\theta\sin\phi,\,\,\,\,\epsilon_{1y}(k)= \cos\theta\sin\phi,\,\,\,\,\epsilon_{2y}(k)= \cos\phi, \nonumber\\
  k_z &=&  \cos\theta,\,\,\,\,\,\,\,\,\,\,\,\,\,\,\,\,\epsilon_{1z}(k)= -\sin\phi,\,\,\,\,\,\,\,\,\,\,\epsilon_{2z}(k)= 0.
  \label{basis}
\end{eqnarray}


\section{Time-evolution of polarized CMB photons}

In this section, we focus on the linear polarization of CMB (E\,- and B\,- modes) due to
the Compton and photon-neutrino (CNB) scattering in company with primordial
scalar perturbations only. As usual, the CMB radiation transfer in the conformal time
$\eta$ is described by
the multipole moments of temperature (I) and polarization (P)
\begin{eqnarray}
\Delta_{I,P}(\eta,K,\mu)=\sum_{\ell=0}^\infty(2\ell+1)(-i)^{l}\Delta_{I,P}^\ell(\eta,K)P_l(\mu),\nonumber
\end{eqnarray}
where $\mu=\hat{\bf n}\cdot \hat{\mathbf{K}}=\cos\theta$,  the
$\theta$ is angle between the CMB photon direction $\hat{\bf n}={\bf k}/|{\bf k}|$ and the wave-vectors $\mathbf{K}$ of Fourier modes of scalar perturbations, and $P_\ell(\mu)$ is the Legendre polynomial of rank $\ell$. We adopt the following Boltzmann equation obeyed by $\Delta_{I,P}(\eta,K,\mu)$, and expand the primordial scalar perturbations $(S)$ of metric field in Fourier modes characterized by the wave-vector $\mathbf{K}$.
For a given Fourier mode, ones can select the
coordinate system where
$\mathbf{K} \parallel \hat{\mathbf{z}}$ and
$(\hat{\mathbf{e}}_1,\hat{\mathbf{e}}_2)=(\hat{\mathbf{e}}_\theta,
\hat{\mathbf{e}}_\phi)$.
For each plane wave, the scattering can be described as the transport through a plane
parallel medium \cite{chandra,kaiser}, and Boltzmann equations are
\begin{eqnarray}
&&\frac{d}{d\eta}\Delta_I^{(S)} +iK\mu \Delta_I^{(S)}+4[\dot{\psi}-iK\mu \varphi]
=\dot\tau\Big[-\Delta_I^{(S)} +
\Delta_{I}^{0(S)} +i\mu v_b +{1\over 2}P_2(\mu)\,\Pi\Big]  \label{Boltzmann}\\
&&\frac{d}{d\eta}\Delta _{P}^{\pm (S)} +iK\mu \Delta _{P}^{\pm (S)} = \dot\tau\Big[
-\Delta _{P}^{\pm (S)} -{1\over 2} [1-P_2(\mu)]\, \Pi\Big]\mp i\,a(\eta) \,\dot{\kappa}_{\pm}\,\Delta _{P}^{\pm (S)}
\label{Boltzmann1}
\end{eqnarray}
where $\dot\tau\equiv d\tau/d\eta$, the scaling factor $a(\eta)|_{\eta_0}=1$ at the present time $\eta_0$, $\Pi\equiv \Delta_{I}^{2(S)}+\Delta_{P}^{2(S)}+\Delta_{P}^{0(S)}$ and the polarization anisotropy is defined by
\begin{eqnarray}
 \Delta _{P}^{\pm (S)}&=&Q^{(S)}\pm iU^{(S)}.\label{pan}
\end{eqnarray}
In the RHS of Eqs.~(\ref{Boltzmann}) and (\ref{Boltzmann1}), the scattering
parts are determined by the Compton and photon-neutrino
scattering terms in Eq.~(\ref{Bo1}), in particular,
the contribution of photon-neutrino (CNB) scattering to CMB polarization
comes from the $\dot{\kappa}_{\pm}$-terms in Eq.~(\ref{Bo1}).
The temperature anisotropy $\Delta_I^{S}$ depends on
the metric perturbations $\varphi$ and $\psi$ and baryon velocity term $v_b$ in Eq.~(\ref{Boltzmann}). Eq.~(\ref{Boltzmann1}) for the polarization anisotropy can be written as follows,
\begin{equation}\label{Boltzmann2}
    \frac{d}{d\eta}\left[\Delta _{P}^{\pm (S)}\,e^{iK\mu\eta\, \pm\, i\tilde{\kappa}(\eta,\mu)\,+\tilde{\tau}(\eta)}\right]  = -e^{iK\mu\eta\, \pm\, i\tilde{\kappa}(\eta)\,+\tilde{\tau}(\eta)}\left(
{1\over 2} \dot\tau[1-P_2(\mu)] \,\Pi\right),
\end{equation}
where
\begin{equation}\label{opticalk}
\tilde{\kappa}(\eta,\mu)\equiv\int_0^{\eta}\, d\eta\, a(\eta)\,\dot{\kappa}_{\pm}, \quad \tilde{\tau}(\eta)\equiv\int_0^{\eta}\, d\eta\, \dot{\tau}.
\end{equation}
With the initial condition $\Delta _{P}^{\pm (S)}
(0,K,\mu)=0$, the integration of
Eq.~(\ref{Boltzmann2}) along the line of sight  up to the present time $\eta_0$ yields \cite{sz}
\begin{eqnarray}
\Delta _{P}^{\pm (S)}
(\eta_0,K,\mu)&=&{3 \over 4}(1-\mu^2)\int_0^{\eta_0} d\eta\,
e^{ix \mu \pm  i\kappa(\eta) -\tau}\,\dot\tau\,\Pi(\eta,K)\label{Boltzmann3}
\end{eqnarray}
where $x=K(\eta_0 - \eta)$ and
\begin{equation}\label{optik}
\kappa(\eta)=\int_{\eta}^{\eta_0}\, d\eta\, a(\eta)\,\dot{\kappa}_{\pm}(\eta).
\end{equation}
These are analogous to
the optical depth $\tau (\eta)$ with respect to the Compton scattering
\begin{equation}\label{optical}
\dot{\tau}=an_ex_e\sigma_T,\,\,\,\,\,\,\,\tau(\eta)=\int_\eta^{\eta_0}\dot{\tau}(\eta) d\eta,
\end{equation}
where $n_e$ is the electron density, $x_e$ is the ionization
fraction and $\sigma_T$ is the Thomson cross-section.

\section{The B-mode power spectrum of polarized CMB photons}

One can separate
the CMB polarization $\Delta _{P}^{\pm (S)}
(\eta_0,K,\mu)$ into the divergence-free part
(B-mode $\Delta_{B}^{(S)}$) and curl-free part (E-mode $\Delta_{E}^{(S)}$) as
following \cite{hu}
\begin{eqnarray}\label{Emode}
\Delta_{E}^{(S)}(\eta_0,K,\mu)&\equiv&-\frac{1}{2}[\bar{\eth}^{2}\Delta_{P}^{+(S)}(\eta_0,K,\mu)+\eth^{2}\Delta_{P}^{-(S)}(\eta_0,K,\mu)]\\
\label{Bmode}\Delta_{B}^{(S)}(\eta_0,K,\mu)&\equiv&\frac{i}{2}[\bar{\eth}^{2}\Delta_{P}^{+(S)}(\eta_0,K,\mu)-\eth^{2}\Delta_{P}^{-(S)}(\eta_0,K,\mu)]
\end{eqnarray}
where $\eth$ and $\bar{\eth}$ are spin raising and lowering operators
respectively, and one assumes scalar perturbations to be axially symmetric
around ${\bf K}$ so that
\begin{eqnarray}
 \bar{\eth}^{2}\, \Delta_{P}^{\pm (S)}(\eta_0,K,\mu)&=&\partial_{\mu}^{2}[(1-\mu^{2})\,\, \Delta_{P}^{\pm (S)}(\eta_0,K,\mu)],\label{axial}
\end{eqnarray}
where $\partial_{\mu}=\partial/\partial\mu$.	
\comment{
for an arbitrary spherical function $ _{\pm 2}f(\theta,\varphi)$ that satisfies
$\partial_{\varphi}\,_{s}f = im  _{s}f$ can be expressed as
\begin{eqnarray}
 \bar{\eth}^{2} _{2}f(\mu,\varphi)&=&(-\partial_{\mu}+\frac{m}{1-\mu^{2}})^{2}[(1-\mu^{2})\,\, _{2}f(\mu,\varphi)],\nonumber\\
  \eth^{2} _{-2}f(\mu,\varphi)&=&(-\partial_{\mu}-\frac{m}{1-\mu^{2}})^{2}[(1-\mu^{2})\,\, _{-2}f(\mu,\varphi)],\label{axial}
\end{eqnarray}
where $\partial_{\mu}=\partial/\partial\mu$ and by assuming $m=0$,
\begin{eqnarray}
 \bar{\eth}^{2}\, _{2}f(\mu,\varphi)&=&\partial_{\mu}^{2}[(1-\mu^{2})\,\, _{2}f(\mu,\varphi)],\nonumber\\
  \eth^{2}\, _{-2}f(\mu,\varphi)&=&\partial_{\mu}^{2}[(1-\mu^{2})\,\, _{-2}f(\mu,\varphi)].\label{axia2}
\end{eqnarray}
}
From Eqs.~(\ref{Boltzmann3}) and (\ref{axial}), we obtain the E- and B-modes
\begin{eqnarray}
\Delta_{E}^{(S)}(\eta_0,K,\mu)&=&
-\frac{3}{4}\int_{0}^{\eta_{0}}d\eta \,g(\eta)\,\Pi(\eta,K)\partial_{\mu}^{2} \left[(1-\mu^{2})^2e^{ix\mu} \cos{\kappa(\eta)}\right], \label{Emode1}\\
\Delta_{B}^{(S)}(\eta_0,K,\mu)&=&\frac{3}{4}\int_{0}^{\eta_{0}}d\eta \,g(\eta)\,\Pi(\eta,K)\partial_{\mu}^{2} \left[ (1-\mu^{2})^2e^{ix\mu} \sin{\kappa(\eta)}\right],\label{Bmode1}
\end{eqnarray}
where $g(\eta)=\dot{\tau}e^{-\tau}$. Eq.~(\ref{Bmode1}) shows that the
photon-neutrino scattering ($\kappa \not=0$) results in the nontrivial
B-mode $\Delta_B^{(S)}$ and the modifications
of the E-mode $\Delta_E^{(S)}$. This agrees that the
Compton scattering only can not generate B-mode without taking into account the tensor type of metric perturbations \cite{zh95,uros,letter,hu}.

Using Eq.~(\ref{Boltzmann3}), we can
obtain the value of $\Delta _{E,B}^{(S)}(\hat{\bf n})$ at the
present time $\eta_0$ and in the direction $\bf \hat{n}$ by summing over all their Fourier modes $K$, analogously to the normal approach \cite{zal,hu,sz},
\begin{eqnarray}
\Delta _{E,B}^{(S)}(\hat{\bf{n}})
&=&\int d^3 {\bf K} \,\xi({\bf K})e^{\mp2i\phi_{K,n}}\Delta _{E,B}^{(S)}
(\eta_0,K,\mu),\label{sumf}
\end{eqnarray}
where
$\phi_{K,n}$ is the angle needed to rotate
the $\bf{K}$ and $\hat{\bf{n}}$ dependent basis to a
fixed frame in the sky. The random variable $\xi(\bf{K})$
used to characterize the initial amplitude of the mode satisfies [see for
example \cite{zal,hu,sz}]
\begin{equation}
\langle \xi^{*}({\bf K}_1)\xi({\bf K}_2)
\rangle=
P_{S}({\bf K})\delta({\bf K}_1- {\bf K}_2),\label{sps}
\end{equation}
where $P_S(K)$ is the initial power spectrum of the scalar mode perturbation.

\begin{figure}
  \includegraphics[width=4in]{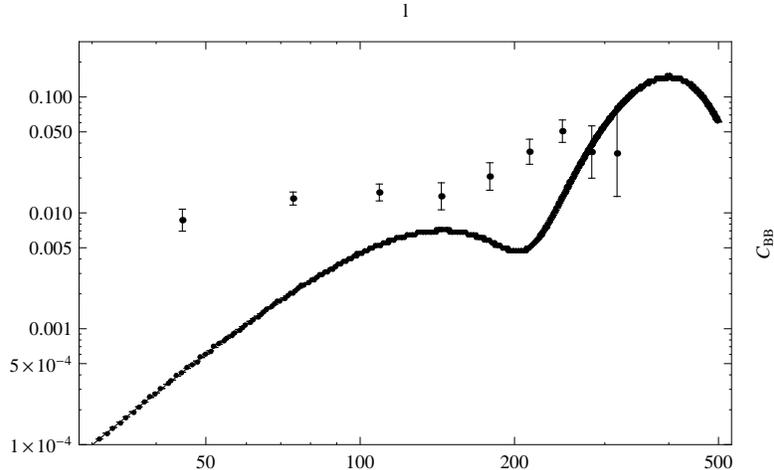}\\
  \caption{The solid line represents $\ell(\ell+1)C_{B}^{\ell(S)}/2\pi[\mu K^2]$ due to the primordial scalar perturbations and photon-neutrino (CNB) scattering. The experiment BICEP2 results (dots with their error bars) are plotted.}\label{fbicep0}
\end{figure}

As a result, by integrating Eqs.~(\ref{sumf}) and (\ref{sps})
over the initial power spectrum of the metric perturbation, we obtain
the power spectrum for $E$- and $B$- modes
\begin{equation}\label{Cl1}
C_{E,B}^{\ell\,(S)}=\frac{1}{2\ell+1}\frac{(\ell-2)!}{(\ell+2)!}\int d^3K P_S(K)\Big|\sum_m\int d\Omega Y^*_{lm}\Delta_{E,B}^{(S)}(\eta_0,K,\mu)\Big|^2.
\end{equation}
Using identities $\partial_{\mu}^{2} (1-\mu^{2})^2e^{ix\mu}\equiv(1+\partial_{x}^{2})x^2\,e^{ix\mu}$ and $\int{d\Omega}\,Y^*_{\ell m}\,e^{ix\mu}=(i)^\ell\,\sqrt{4\pi(2\ell+1)}\,j_\ell(x)\delta_{m0}$, we obtain the polarized CMB power spectrum in multipole moments $\ell$,
\begin{eqnarray}
  C_{E}^{ \ell(S)}&=&(4\pi)^2\frac{(\ell+2)!}{(\ell-2)!}\int d^3K P_S(K)\Big|\frac{3}{4}\int_0^{\eta_0}d\eta\,g(\eta)\,\Pi(\eta,K)\frac{j_\ell}{x^2} \cos{\kappa(\eta)}\Big|^2,\label{Emode2} \\
  C_{B }^{\ell(S)}&=&(4\pi)^2\frac{(\ell+2)!}{(\ell-2)!}\int d^3K P_S(K)\Big|\frac{3}{4}\int_0^{\eta_0}d\eta\,g(\eta)\,\Pi(\eta,K)\frac{j_\ell}{x^2} \sin{\kappa(\eta)}\Big|^2.\label{Bmode2}
\end{eqnarray}
where $j_\ell(x)$ is a spherical Bessel function of rank $\ell$. In
Fig.~\ref{fbicep0}, we plot the numerical value $C_{B }^{\ell(S)}$ of Eq.~(\ref{Bmode2}) together
with the BICEP2 result. It is shown
that the contribution of photon-neutrino (CNB) scattering to the B-mode is
negligible for small $\ell$, however is significantly large for $50 < \ell< 200$.

To order to better understand our results (\ref{Emode2}) and (\ref{Bmode2}), we approximately write $C_{E}^{\ell (S)}$ and $C_{B}^{\ell (S)}$ as follows
\begin{eqnarray}
  C_{E}^{\ell (S)}&\approx &\bar{C}^{\ell  (S)}_{E} \, (\cos^2{\bar{\kappa}}),\,\,\,\,\,\,\,\,\,\,\,C_{B}^{\ell (S)}\approx \bar{C}^{\ell  (S)}_{E} \, (\sin^2{\bar{\kappa}})\label{Emode3}
\end{eqnarray}
where
 \begin{eqnarray}
 \bar{C}^{\ell  (S)}_{E}&=&(4\pi)^2\frac{(l+2)!}{(l-2)!}\int d^3K P_S(K)\Big|\frac{3}{4}\int_0^{\eta_0}d\eta\,g(\eta)\,\Pi(\eta,K)\frac{j_\ell}{x^2} \Big|^2,\label{Emode4}
\end{eqnarray}
is the power spectrum of the E-mode polarization contributed
from the Compton scattering in the case of scalar perturbation \cite{zal}.
In Eq.~(\ref{Emode3}), the mean
value $\bar\kappa$ of Eq.~(\ref{optik}) is an average
from the last scattering time (redshift $z_l\approx 10^3$) to the present time ($z_0=0$). Using the matter dominate Friedmann equation
$H^2/H^2_0=\Omega_M^0(1+z)^{3}+\Omega_\Lambda^0, \, H_0\approx 74\,$km/s/Mpc,
$\Omega_M^0\approx 0.27, \Omega_\Lambda^0\approx 0.73$, and
$ad\eta=-dz/H(1+z)$, as well as the conservation of total neutrino number
$n_\nu=n_\nu^0 (1+z)^3$, we obtain
\begin{eqnarray}
\kappa(z) &= &\int^{\eta_{0}}_{\eta}a\,d\eta \dot{\kappa}_{\pm}
=\frac{\sqrt{2}}{12\pi}\alpha\,G^Fn^0_{\nu}\int^{z_{\rm l}}_{z}dz'
\frac{(1+z')^2}{H(z')} \nonumber\\
&=&\frac{\sqrt{2}}{12\pi}\alpha\,G^Fn^0_{\nu}\frac{2H(z')}{3\Omega_M^0H_0^{2}}\Big|_{z}^{z_{l}}\nonumber\\
\bar\kappa &\equiv &\frac{1}{z_{\rm l}-z_0}\int_{z_{\rm l}}^{z_0}dz\,\kappa(z)\approx 0.16,\label{barkappa}
\end{eqnarray}
where the present number-density of all flavor neutrinos and anti-neutrinos
$n^0_{\nu}=\sum(n^0_\nu+n^0_{\bar{\nu}})\approx 340\, {\rm cm^{-3}}$).
Actually, the $\bar\kappa$ is the mean opacity of CMB photons against the
photon-neutrino (CNB) scattering.

To end this section, we would like to point out that the B-mode power spectrum
$C_{B}^{\ell (S)}$ of Eq.(\ref{Bmode2}) is attributed only to the scalar
perturbations and photon-neutrino (CNB) scatterings. This result implies that
the B-mode power spectrum could be contributed by other mechanisms with
scalar perturbations, in addition to the contribution from the primordial
tensor perturbations \cite{zal}. Therefore, this is crucial how to interpret
the measurement of $r$-parameter that is the ratio of the B-mode power
spectrum and the E-mode power spectrum.

\comment{Whereas the mean value $\bar{\kappa}$ is given by the time averaged $\kappa(\eta)$ of Eq.~(\ref{optik}) from the last scattering to the
present time
\begin{equation}\label{kappa-bar}
    \bar{\kappa}\simeq\frac{\sqrt{2}}{12\pi}\alpha\,G^F\,\bar{n}_{\nu}\int_{t}^{t_0}\, dt\simeq\frac{\sqrt{2}}{12\pi}\alpha\,G^F\,\bar{n}_{\nu}H_0^{-1}\sim 0.15,
\end{equation}
where $\bar{n}_{\nu}=\sum_j(\bar{n}_\nu+\bar{n}_{\bar{\nu}})\sim10^8 {\rm cm^{-3}}$ is average CNB number density over time interacting and $H_0$ is Hubble parameter.
\begin{equation}
\kappa=\int_{\eta}^{\eta_0}\, d\eta a\,\dot{\kappa}_{\pm}
=\frac{\sqrt{2}}{12\pi}\alpha\,G^F\,n^0_{\nu}H_0^{-1}\int^{z_{lss}}_{z}\, (dz/H)\,(1+z)^2\sim
\int_{a_l}^{a_0}\, da a\,n_\nu/H=....\nonumber
\end{equation}}
\section{ Summary.}
Suppose that the total contribution to the the polarized CMB B-mode
comes from the primordial tensor perturbations ($T$),  one obtains the B-mode power spectrum \cite{zal}
\begin{eqnarray}
C_{B }^{\ell (T)}&=&(4\pi)^2\int d^3K P_T(K)\Big|\frac{3}{4}\int_0^{\eta_0}d\eta\,g(\eta)\,S^T(\eta, K)\,(2j\,'_{\ell}+\frac{j_\ell}{x}) \Big|^2,\label{Emode2'} \\
 S^T(\eta, K)&=&\left[\frac{1}{10}\Delta^{0(T)}_{I}+\frac{1}{7}\Delta^{2(T)}_{I}+\frac{3}{70}\Delta^{4(T)}_{I}-\frac{3}{5}\Delta^{0(T)}_{P}
 +\frac{6}{7}\Delta^{2(T)}_{P}-\frac{3}{70}\Delta^{4(T)}_{P}\right],\label{Bmode2'}
\end{eqnarray}
where $P_T(K)$ is the initial power spectrum of primordial tensor
perturbations. Based on this assumption, one can approximately obtain
the $r$-parameter $r=P_T/P_S\propto C^{\ell (T)}_{B}/\bar{C}^{\ell (S)}_{E}$.
Taking into account the contribution (\ref{Emode2}) or
(\ref{Emode3}) of photon-neutrino (CNB) scattering and assuming the total observed B-mode power spectrum $C^{\ell (ob)}_{B}$
given by $C^{\ell (ob)}_{B}=C^{\ell (T)}_{B}+C^{\ell (S)}_{B}$, we have
\begin{equation}\label{rr}
r=P_T/P_S\propto (C^{\ell (ob)}_{B}-C^{\ell (S)}_{B})/ \bar{C}^{\ell (S)}_{E}\approx C^{\ell (T)}_{B}/ \bar C^{\ell (S)}_{E}-\sin^2{\bar{\kappa}},
\end{equation}
where $\sin^2{\bar{\kappa}}\sim\bar{\kappa}^2\simeq 0.025$.  This implies that the measured $r$-parameter would not be completely originated from primordial tensor
perturbations. In addition, there might be  other contributions from
either some astrophysical effects \cite{Bovin,Giovannini} or
some microscopic effects, for example the CMB photon-photon scatterings \cite{xue,EH}.
Therefore, it is important to study possibly significant contributions to the
B-mode power spectrum of polarized CMB photons so that one can better
understand the contribution of primordial tensor perturbations to the $r$-
parameter experimentally measured, $r=0.2$ as reported by BICEP2.

In summary, we have studied the Quantum
Boltzmann Equation governing the time-evolution of the density matrix (Stokes parameters) of polarized CMB photons by considering the
both Compton and photon-neutrino (CNB) scattering in the background of
primordial scalar perturbations. It is shown that in this case
the B-mode spectrum of polarized CMB photons can also be generated without
primordial tensor perturbations. We quantitatively calculate the
generated B-mode spectrum which is related to the mean opacity $\bar\kappa$ (\ref{barkappa}) of CMB photons scattering with neutrinos (CNB). In the other hand, we compare our
result with the B-mode spectrum generated by the Compton scattering in the
background of primordial tensor perturbations, which seems to be dominated.
We generally discuss the possible implication of our result
on the interpretation of the BICEP2 measurement $r=0.2$ in terms of
primordial tensor perturbations.


\end{document}